\documentclass[prl,twocolumn,showpacs,floatfix]{revtex4}
\usepackage{graphics}
\usepackage{graphicx}
\begin{document}
\newcommand{\RR}{\mathrm{\mathbf{R}}}
\newcommand{\rr}{\mathrm{\mathbf{r}}}
\newcommand{\defin}{\stackrel{def}{=}}

\title{Magnetoresistance in an all-manganite heterostructure}
\author{J. Salafranca}
\affiliation{Instituto de Ciencia de Materiales de Madrid (CSIC), Cantoblanco, 28049 Madrid (Spain)}
\author{M.J. Calder\'on}
\affiliation{Instituto de Ciencia de Materiales de Madrid (CSIC), Cantoblanco, 28049 Madrid (Spain)}
\author{L. Brey}
\affiliation{Instituto de Ciencia de Materiales de Madrid (CSIC), Cantoblanco, 28049 Madrid (Spain)}

\date{\today}

%$\lesssim$

%$\gtrsim$

\begin{abstract}
We study the magnetic and transport properties of all-manganite heterostructures consisting of ferromagnetic metallic electrodes separated by an antiferromagnetic barrier. We find that the magnetic ordering in the barrier is influenced by the relative orientation of the electrodes magnetization producing a large difference in resistance between the parallel and antiparallel orientations of the ferromagnetic layers. The external application of a magnetic field in a parallel configuration also leads to large magnetoresistance.
\end{abstract}

\pacs{75.47.Gk, 75.10.-b, 75.30.Kz, 75.50.Ee
}
\maketitle
%Motivation:experiments
%Strongly correlated system heterostructures (Millis): electronic reconstruction
Heterostructures of strongly correlated systems are currently in the spotlight due to the 
appearance of new electronic phases (electronic reconstruction)~\cite{okamoto-nat04,huijben06,lin06,kancharla06,brey-PRB07} at interfaces between materials with strong electron-electron and electron-lattice interactions. Manganese perovskites are specially interesting because of their potential application in spintronics~\cite{zuticRMP}: in the ferromagnetic phase they are half-metals~\cite{pickett96,park98} and, therefore, very efficient spin injectors and detectors~\cite{bowen03,yamada04,bibes-review}. In bulk or thin films, manganites show an extremely large (colossal) magnetoresistance, and phase separation, fruit of the competition between very different phases ranging from metallic and ferromagnetic (FM) to insulating and antiferromagnetic (AF)~\cite{dagotto-book,israel07}. The different phases arise as a function of doping and composition. For instance, bulk La$_{1-x}$Sr$_{x}$MnO$_3$ (LSMO) is FM and metallic at the so-called optimal doping ($x \sim 1/3$) with a $T_C$ above room temperature, while systems with smaller A-site cations, like bulk Pr$_{1-x}$Ca$_{x}$MnO$_3$ (PCMO), are AF (CE-type ordering~\cite{brink99}) and insulating (with charge/orbital order) at the same doping. 

Manganite surfaces are known to behave differently from the bulk, the typical example being the striking suppression of the spin polarization of a free surface at temperatures much lower than the bulk ferromagnetic $T_C$~\cite{park98b,calderonsurf}. This could have a very negative effect on the tunneling magnetoresistance (TMR)~\cite{julliere75} of magnetic tunnel junctions consisting of half-metallic manganites separated by a thin insulating layer because TMR depends very strongly on the properties of the electrode/barrier interface~\cite{leclair00}. Indeed, early reports of TMR in manganite heterostructures showed a very strong decrease with increasing temperature~\cite{lu96}. TMR is defined as the difference in resistance $R$ between parallel (P) and antiparallel (AP) relative orientations of the magnetization in the ferromagnetic metallic electrodes [TMR$=(R_{\rm AP}-R_{\rm P})/R_{\rm AP}$]. Recently it has been found that interfaces of manganites with suitable materials are able to keep the spin polarization close to that of the bulk up to higher temperatures~\cite{yamada04,garcia04,ishii06} with the concomitant enhancement of TMR. Traditionally, insulating barriers such as SrTiO$_3$, LaAlO$_3$, and NdGaO$_3$ are used.

\begin{figure}
%\begin{center}
\resizebox{65mm}{!}{\includegraphics{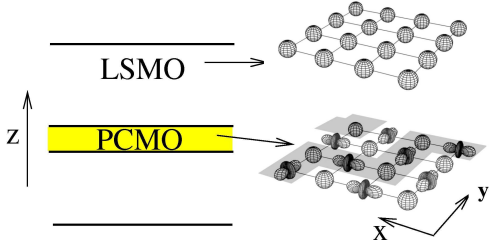}}
\caption{\label{fig:scheme} Schematic view of the heterostructure under consideration. LSMO stands for La$_{2/3}$Sr$_{1/3}$MnO$_3$ and PCMO for Pr$_{2/3}$Ca$_{1/3}$MnO$_3$. At this doping ($x=1/3$), bulk LSMO is FM and metallic, and bulk PCMO is CE-type AF, with FM zig-zag chains in the xy-plane antiferromagnetically coupled to neighboring chains, orbital/charge-ordered and insulating. 
}
%\end{center}
\end{figure}

%What we study and what we find
Here we focus on all-manganite heterostructures where the barriers are AF insulating manganites and study the effect of the FM layers on the ground state configuration at the barrier. These barriers are qualitatively different from other insulating barriers in that their electric and magnetic properties are expected to be influenced by the FM electrodes. 
In particular, we study the trilayer La$_{2/3}$Sr$_{1/3}$MnO$_3$/Pr$_{2/3}$Ca$_{1/3}$MnO$_3$/La$_{2/3}$Sr$_{1/3}$MnO$_3$ illustrated in Fig.~\ref{fig:scheme} ~\cite{li02,niebieskikwiat07}. 
The LSMO layers are considered to have in-plane FM polarization (x-direction) in a P or AP configuration. 

\begin{figure}
%\begin{center}
\resizebox{58mm}{!}{\includegraphics{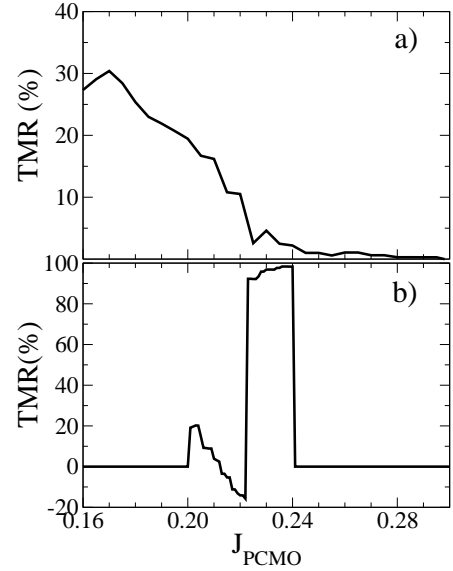}}
%\resizebox{70mm}{!}{\includegraphics{TMRvs_JPC_3.pdf}}
\caption{\label{fig:TMRvsJ} Tunneling magnetoresistance versus $J_{\rm PCMO}$ calculated for PCMO layer thicknesses of two (a) and three (b) lattice parameters. For large values of $J_{\rm PCMO}$, the TMR is very small because the PCMO spacer is AF and insulating for both P and AP configurations. For $J_{\rm PCMO} < 0.24 \, t$ (a) and (b) show different qualitative behaviors (see text for discussion). The maximum sensitivity to magnetization is reached in PCMO-3 for $0.22 \, t < J_{\rm PCMO} < 0.24 \, t$ where the system is metallic in the P configuration while insulating in the AP configuration. 
}
%\end{center}
\end{figure}

\begin{figure}
%\begin{center}
\resizebox{60mm}{!}{\includegraphics{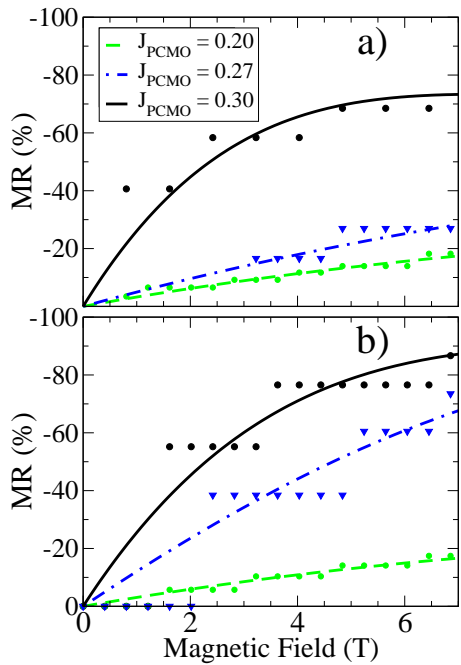}}
\caption{\label{fig:MRvsH} (Color online) Magnetoresistance in the parallel configuration upon application of a small magnetic field in the x-direction for three different values of $J_{\rm PCMO}$ (a) PCMO-2 and (b) PCMO-3. $t=0.25$ eV is used for the estimation of the magnetic field $H$. The lines are fits to the dots. 
}
%\end{center}
\end{figure}

We find that the ground state configuration in the PCMO layer depends on the relative orientation of the magnetization in the LSMO layers and, as a consequence, the system shows a large TMR (see Fig.~\ref{fig:TMRvsJ}). In general terms, the P configuration induces FM ordering in the PCMO layer while the AP configuration tends to stabilize AF ordering in the barrier. The influence of the FM electrodes on the PCMO layer is due to a balance between the different coupling strengths in LSMO and PCMO and the charge transfer between the different layers.
In principle, although the doping of divalent atoms $x$ is constant throughout the system, the strong electron-lattice interaction in the PCMO layer opens a gap in the density of states which produces a transfer of electrons towards the metallic FM areas. However, this charge transfer is limited by the long-ranged Coulomb interaction and, consequently, the PCMO layer can get more conductive and less AF than in the bulk~\cite{footnote-insulating}.   
We also show that, in the P configuration, the application of a parallel magnetic field (along the x-direction) affects the PCMO layer magnetic ordering giving rise to negative MR (see Fig.~\ref{fig:MRvsH}).

%The Model
We address these issues by finding the minimal energy spin, charge and orbital configuration in a very thin PCMO spacer [two (PCMO-2) to three (PCMO-3) lattice parameters, $a$, wide] between two wider and perfectly ferromagnetic LSMO layers.
The tight-binding Hamiltonian has the following terms\cite{brey-PRB07}:
\begin{eqnarray}
H &=&- \sum_{i,j,\gamma,\gamma'} f_{i,j} t^u_{\gamma,\gamma'} C_{i,\gamma}^{\dagger} C_{j,\gamma'} + \sum_{i,j}  J_{AF}^{ij} {\mathbf S}_i  {\mathbf S}_j  \nonumber \\ 
&+& U'\sum_{i} \sum_{\gamma \ne \gamma'} n_{i \gamma} n_{i \gamma'}+ H_{\rm Coul}
\label{eq:H}  
\end{eqnarray}
where $C_{i,\gamma}^{\dagger}$ creates an electron on the Mn i-site, in the $e_g$ orbital $\gamma$ ($\gamma=1,2$ with $1=|x^2-y^2 \rangle$ and $2=|3 z^2-r^2 \rangle$). $\langle n_i \rangle=\sum_{\gamma} \langle  C_{i,\gamma}^{\dagger} C_{i,\gamma} \rangle$ is the occupation number on the Mn i-site. The hopping amplitude depends on the Mn core spins orientation given by the angles $\theta$ and $\psi$ via $f_{i,j}=\cos(\theta_i/2) \cos(\theta_j/2)+ \exp[i(\psi_i-\psi_j)]\sin(\theta_i/2) \sin(\theta_j/2) $ (double-exchange mechanism), and on the orbitals involved $t^{\rm x(y)}_{1,1}=\pm \sqrt{3} \,t^{\rm x(y)}_{1,2}=\pm \sqrt{3}\, t^{\rm x(y)}_{2,1}=3\, t^{\rm x(y)}_{2,2}=3/4 \,t^{\rm z}_{2,2}=t$ where the superindices x,y, and z refer to the direction in the lattice. All the parameters are given in units of $t$ which is estimated to be $\sim 0.2 - 0.5$ eV. $J_{AF}$ is an effective antiferromagnetic coupling between first neighbor Mn core spins which is different in the LSMO and PCMO layers (see below).
$U'$ is a repulsive interaction between electrons on a site lying on different orbitals,  and $H_{\rm Coul}$ is the long range Coulomb interaction between all the charges in the system, treated in the mean-field approximation
{\small
\begin{equation}
H_{\rm Coul} ={{\frac{e^2}{\epsilon}}} \sum_{i \ne j} \left({\frac{1}{2}} {\frac {\langle n_i \rangle   \langle n_j \rangle}{|{\mathbf R}_i-{\mathbf R}_j |}} +{\frac{1}{2}} {\frac {Z_i Z_j}{|{\mathbf R}^A_i-{\mathbf R}^A_j |}}
-{\frac{Z_i   \langle n_j \rangle}{|{\mathbf R}^A_i-{\mathbf R}_j |}}\right)
\end{equation}
}%
with ${\mathbf R}_i$ the position of the Mn ions, $eZ_i$ the charge of the A-cation located at ${\mathbf R}_i^A$, and $\epsilon$ the dielectric constant of the material. The strength of the Coulomb interaction is given by the dimensionless parameter $\alpha=e^2/a \epsilon t$~\cite{lin06}. 

The electron-lattice interaction has not been explicitely included in the Hamiltonian~(\ref{eq:H}). However, the effect of this coupling on the ground state energies can be described using an effective $J_{AF}$~\cite{vandenbrink-PRL99}. In particular, the ground state of Hamiltonian~(\ref{eq:H}) for a bulk system with  $J_{AF} \gtrsim 0.2 t$ is the CE-type AF ordering associated to the lattice distortions that produce the charge and orbital ordering illustrated in Fig.~\ref{fig:scheme}.
The values for $J_{AF}$ that effectively include the electron-lattice coupling are therefore larger than the ones inferred from the magnetic ordering only ($J_{AF}^S$ from superexchange between the $t_{2g}$ electrons is $\sim 1-10$ meV)~\cite{dagotto-book}. 
The FM layers (LSMO) are well 
described by small values of $J_{AF}$ (for simplicity, we consider $J_{\rm LSMO}=0$), while for the barrier (PCMO) 
we use $0.15 \, t < J_{\rm PCMO} < 0.3 \, t$. Reasonable values for the other two parameters are $U'=2 t$ and $\alpha=2$~\cite{brey-PRB07}. The results presented below are qualitatively insensitive to moderate changes of these two parameters. The Hamiltonian is solved self-consistently in heterostructures consisting of a thin  PCMO layer and two wide LSMO FM layers (wide enough to reproduce bulk behavior).

\begin{figure}
%\begin{center}
\resizebox{59mm}{!}{\includegraphics{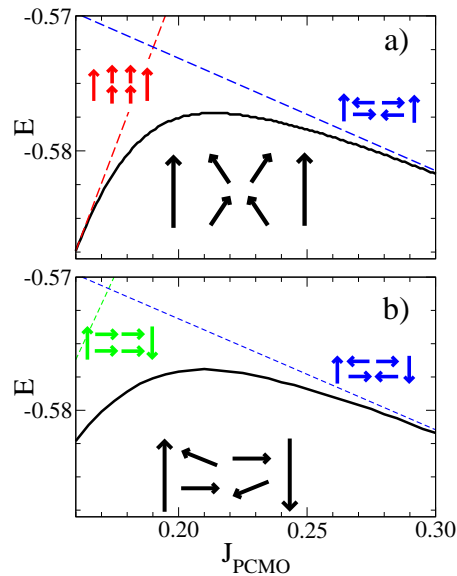}}
\caption{\label{fig:energyvsJ} 
(Color online) Energy versus $J_{\rm PCMO}$ for a parallel (a) and antiparallel (b) configuration of the LSMO layers, with PCMO thickness of two lattice parameters, PCMO-2. The dashed lines are the energy for the pure FM and pure CE configurations in the intermediate PCMO layer.  The actual ground state (solid line) corresponds to canted intermediate configurations (illustrated in the insets). The big arrows represent the magnetization orientation in the FM layers and the small ones represent the order considered in the PCMO-2 layer. In the CE and canted phases each arrow in the PCMO layer represents a FM zig-zag chain (see Fig.~\ref{fig:scheme}).
}
%\end{center}
\end{figure}

%Results and discussion: energy, ground state configuration
In Fig.~\ref{fig:energyvsJ} we show the total energy versus $J_{\rm PCMO}$ for a pure FM, CE and an intermediate {\em canted} configuration of the PCMO-2 layer for P and AP configurations of the electrodes.  All the other magnetic orderings considered were higher in energy in this range of parameters.  For $0.17 \, t < J_{\rm PCMO} < 0.3 \, t$, the magnetic ground state configuration in the spacer is always canted with the canting angle depending on the value of $J_{\rm PCMO}$ and on the relative orientation of the magnetization in the LSMO layers. In the P configuration [Fig.~\ref{fig:energyvsJ}(a)], PCMO tends to order more FM and collinearly with the electrodes, while for the AP case [Fig.~\ref{fig:energyvsJ}(b)], the PCMO configuration corresponds to smaller magnetization and the spins lie perpendicular to the electrodes magnetization. The results for PCMO-3 (not shown) are qualitatively similar. 

%TMR
For PCMO-2 and $J_{\rm PCMO}$ relatively small ($ \lesssim 0.24 \, t$), the magnetic order at the barrier is canted, and
the charge/orbital order is mostly suppressed due to
charge transfer between the layers. 
FM correlations and, due to double exchange, conductance are larger in the P configuration than in the AP configuration.
Therefore, this geometry could be used as a magnetic sensor. The conductance has been calculated numerically via Kubo formula~\cite{verges,cond} for a trilayer with semi-infinite FM LSMO leads. The results for a PCMO-2 spacer are plotted in Fig.~\ref{fig:TMRvsJ} (a) where a finite TMR for $J_{\rm PCMO} \lesssim 0.24 \, t$ is shown. The superstructure in the curve is due to numerical inaccuracies except for the peak at $J_{\rm PCMO} \sim 0.17 \, t$, which is quite robust (the TMR increases monotonically in the range $0.1 \, t < J_{\rm PCMO} \lesssim 0.17 \, t$). This peak appears because, below $\sim 0.17 \, t$, the P ground state configuration is almost FM [Fig.~\ref{fig:energyvsJ} (a)] while the AP configuration is already canted [Fig.~\ref{fig:energyvsJ} (b)] and, as a consequence, $R_{\rm AP}$ increases faster with $J_{\rm PCMO}$ than $R_{\rm P}$.  

For PCMO-3 there is a range of parameters, $0.22 \, t \le J_{\rm PCMO} \le 0.24 \, t$, for which the TMR is close to its maximum possible value of $100\%$. In this range, the P configuration is metallic as it has a relatively large FM component in the three Mn planes that constitute the barrier, while the AP configuration is insulating and corresponds to perfect CE in the middle atomic plane and canted FM in the outer planes. 
For smaller values of $J_{\rm PCMO}$  ($\le 0.2 \, t$), for both the P and AP configurations, the middle plane is a perfect CE while the outer planes are essentially FM and parallel to the nearest electrode; this leads to $R_{\rm P}= R_{\rm AP}$ and, hence, TMR$=0$. The negative TMR at $J_{\rm PCMO} \sim 0.22 \, t$ is produced by the different dependence of the canting angle on  $J_{\rm PCMO}$  for P and AP configurations. The different behavior of the TMR in PCMO-2 and PCMO-3 is due to the limited charge transfer in the middle Mn plane of the wider barrier.

For large $J_{\rm PCMO}$ ($>0.24 \, t$), see Figs.~\ref{fig:TMRvsJ} (a) and (b),  or wider PCMO spacers (PCMO-n with $n \ge 4$), the AF ordering in the barrier is preserved and the system does not show TMR at all: both $R_{\rm P}$ and $R_{\rm AP}$ are strictly $0$.

%MR
We have also calculated the MR in the P configuration that results of applying an external magnetic field parallel to the magnetization in the electrodes. The results are shown for PCMO-2 and PCMO-3 in Fig.~\ref{fig:MRvsH}. We define MR$=(R(H)-R(0))/R(0) \times 100\%$ so its maximum possible absolute value is MR$=100\%$. The dots represent the numerical values and the steps are an artifice of the calculation that considers a discrete set of values of the canting angle. The lines are a fit to the data. When a magnetic field $H$ is applied, the system is effectively moving towards smaller values of $J_{\rm PCMO}$ (see Fig.~\ref{fig:energyvsJ}) and therefore towards less resistive configurations, hence the negative MR. This MR is produced by the alignment of the barrier spins with the applied field and is smaller than the CMR measured in bulk PCMO~\cite{anane99} which is probably related to inhomogeneities and phase separation. The real advantage of this heterostructure as a magnetic sensor is that its resistivity can be orders of magnitude smaller than the bulk PCMO's (mainly because there is no gap at the Fermi energy for the thin spacers in the P configuration). As a guideline, the resistivity of bulk LSMO at low $T$ is $\sim 10^{-4} \, \Omega.$cm~\cite{urushibara95}, much smaller than that of bulk PCMO $\gtrsim 10^{5} \,\Omega.$cm ($\sim 10^{-3} \,\Omega.$cm at $7$ T)~\cite{yoshizawa96}.

It is well known that strain (produced by lattice mismatch between the substrate and the thin films) can affect the orbital ordering~\cite{tokura00}. In the studied heterostructure with an STO substrate~\cite{niebieskikwiat07} the in-plane lattice parameter is $3.90 {\rm \AA}$ for all layers while the out-of-plane lattice parameters are $3.85 {\rm \AA}$ (LSMO) and $3.76{\rm \AA}$ (PCMO), slightly smaller (less than a $2 \%$ in any case) than the bulk values. Our calculations are done in a cubic lattice but the variations in unit cell dimensions in actual heterostructures~\cite{niebieskikwiat07} are not expected to produce a dramatical change in the orbital ordering~\cite{tokura00}. In any case it would emphasize the tendency to CE ordering in the PCMO barrier that can be included in our model simply by increasing the value of $J_{\rm PCMO}$. Strain can also produce phase separation~\cite{fontcuberta-preprint} and  
colossal magnetoresistance~\cite{ahn04}. The inclusion of phase separation in our model would lead to an increase of both TMR (Fig.~\ref{fig:TMRvsJ}) and MR (Fig.~\ref{fig:MRvsH}) with respect to the calculated values.

%Summary and conclusion
In conclusion, we have studied the electric and magnetic properties of an all-manganite heterostructure with homogeneous doping of divalent cations. The system consists of two FM and metallic electrodes (La$_{2/3}$Sr$_{1/3}$MnO$_3$) and a thin AF barrier (Pr$_{2/3}$Ca$_{1/3}$MnO$_3$). We show that the ground state configuration in the PCMO layer depends on the relative orientation of the FM electrodes (parallel or antiparallel) rendering these heterostructures useful as magnetic sensors. Underlying this result is the fact that charge transfer between the layers with different coupling strengths alters the electric and magnetic properties of the thin PCMO layer (which is AF and insulating in bulk) increasing its magnetization and conductivity. Therefore, the resulting heterostructures have low resistivity in a wide range of parameters, mainly in the parallel configuration, at which an external field applied can produce a large negative MR.

We thank J.A. Verg\'es for fruitful discussions. This work is supported by MAT2006-03741 (MEC, Spain). M.J.C. also acknowledges Programa Ram\'on y Cajal (MEC, Spain).

\bibliography{manganites}

\begin{thebibliography}{33}
\bibitem[{foo()}]{footnote-insulating}
\bibinfo{note}{Note that PCMO at $x=0.5$ is insulating because there is a gap
  at the Fermi energy, while for $x<0.5$, which is the case here, PCMO is
  insulating when AF ordering produces no extended states at the Fermi energy.}
\end{thebibliography}

\end{document}